# CPT Symmetry in Projective de Sitter Universes

Ignazio Licata[1,2], Davide Fiscaletti[3], Leonardo Chiatti[4] and Fabrizio Tamburini[5]

[1] ISEM, Institute for Scientific Methodology, Palermo, Italy.
[2] School of Advanced International Studies on Applied Theoretical and Non-Linear Methodologies in Physics, Bari, Italy.
[3] SpaceLife Institute, San Lorenzo in Campo, PU 61047, Italy.
[4] AUSL VT Medical Physics Laboratory, Via Enrico Fermi 15, 01100 Viterbo, Italy.
[5] ZKM Karlsruhe, Lorentzstraße 19, 76135, Karlsruhe, Germany.

Corresponding Authir Ignazio Licata
E-mail: ignazio.licata3@gmail.com

## Abstract

In a recent work, Boyle, Finn and Turok hypothesized a model of universe that does not violate the CPT-symmetry as alternative for inflation. With this approach they described the birth of the Universe from a pair of universes, one the CPT image of the other, living in pre- and post-big bang epochs. The CPT-invariance strictly constrains the vacuum states of the quantized fields, with notable consequences on the cosmological scenarios. Here we examine the validity of this proposal by adopting the point of view of archaic cosmology, based on de Sitter projective relativity, with an event-based reading of quantum mechanics, which is a consequence of the relationship between the universal information reservoir of the archaic universe and its out-of-equilibrium state through quantum jumps. In this scenario, the big bang is caused by the instability of the original (pre)vacuum with respect to the nucleation of micro-events that represent the actual creation of particles. Finally, we compare our results with those by Boyle *et al*., including the analytic continuation across the big bang investigated by Volovik and show that many aspects of these cosmological scenarios find a clear physical interpretation by using our approach. Moreover, in the archaic universe framework we do not have to assume a priori the CPT-invariance like in the other models of universe, it is instead a necessary consequence of the archaic vacuum structure and the nucleation process, divided into two specular universes.

## 1. Section heading

In the recent paper [1], in order to explain the fact that immediately after the big bang the Universe can be associated to a spatially flat, radiation dominated FRWL metric, Boyle, Finn and Turok have considered a kind of cosmology dictated by a CPT symmetry. This approach, presented as an alternative to the inflation, assimilates the Universe to a CPT invariant vacuum. The Universe after the big bang is the CPT image of the Universe before it and the pre- and post-big bang epochs form a universe/anti-universe pair, emerging from nothing directly into a hot, radiation dominated era. The request of CPT symmetry selects a unique quantum field theory (QFT) vacuum state, providing a new interpretation of the cosmological baryon asymmetry as well as a remarkably economical explanation for the cosmological dark matter. In this way, the observed oscillations characterizing the CMB power spectrum, which in the standard view are usually attributed to inflation, are instead alternatively reproduced with the correct phases by a symmetry argument.

In Ref. [2] Volovik considered two different ways of analytic continuation across the big bang of the model proposed in [1]. On one hand, by extending the conformal time to the temperature. In this picture, the initial stage of the after-big bang evolution is characterized by negative temperatures. On the other hand, the analytic continuation in the proper time leads to the sign change of the scale factor around the big bang; the spacetime thus becomes an anti-spacetime. In this last scenario, the big bang emerges as the bifurcation point of the second-order quantum transition from the Euclidean to the Minkowski signature, at which the symmetry between the spacetime and anti-spacetime is spontaneously broken (and their quantum superposition is no more allowed).

In our opinion, the crucial aspect of Boyle, Finn and Turok (BFT) model lies in the fact that, by imposing a CPT

symmetry to the Universe, one constrains the vacuum. In this context one can perceive a significant physical parallelism to what occurs in our approach of the so-called Archaic Universe [3,4], where a native CPT symmetry is present [5].

Let us briefly recall the concept of "Archaic Universe". This model of universe takes its origins from a cosmological model based on a Euclidean timeless spatial structure. The simplest geometric structure is a Euclidean 5-sphere that, through projective relativity techniques, becomes a substratum of the ordinary spacetime metrics in which the wavefunctions of any quantum state can propagate. The hyper-sphere is like a highly non-local temporal archaic phase. In this geometrical structure, the time variable represents a cosmic time and justifies the assumption of a cosmological principle, identifying the class of the de Sitter observers, a substratum of fundamental observers defined by the big bang. Through a Wick rotation and the emergence of the arrow of time, the transition from a non-local "archaic" phase to a local one occurs, in a quantum interpretation of the big bang. Two well-known difficulties of the standard models of cosmology are worked around in this way, namely, the definition of the initial homogeneity and the singularity problem. Homogeneity and isotropy of spacetime present in the laws of physics are a consequence of the archaic vacuum.

In this work we will show how the CPT symmetry is intrinsically given and embedded in this approach to cosmology, instead of being postulated ad hoc as it happens in the BFT model. Moreover, we will provide a suggestive explanation that the big bang emerges as the bifurcation point of the second-order quantum transition from the Euclidean to the Minkowski signature, at which the symmetry between the spacetime and anti-spacetime is spontaneously broken, as claimed by Volovik.

The idea of an Archaic Universe underlies an objective interpretation of micro-events at the quantum level. This interpretation was outlined in a series of previous works that can be find in the literature and a recent general introduction is reported in Ref. [6]. Here we use a slightly modified version of the micro-event model illustrated in [7]. In this model is assumed a symmetry breaking, which leads from a timeless vacuum to the ordinary vacuum in which micro-events "occur" in a temporal succession. We identify the intermediate phases of this passage with the pre-vacuum previously hypothesized in the model of the Archaic Universe. This identification replaces the analytical continuation procedure studied by Volovik, leading to a thermal time without negative temperatures and to a pre-big bang phase without variation in the sign of the scale factor. This phase is the CPT mirror of the post-big bang phase.

The structure of this paper is the following: in section 2 we recall the fundamental concepts regarding the Archaic Universe and we show how in this approach a CPT symmetry is naturally embedded. In section 3 the comparison with the model described in [1] is discussed. Section 4 addresses the relation between the archaic scenario and the structure of micro-events at the quantum level. Section 5 contains the concluding remarks.

1. **The Archaic Universe in a nutshell**

The fundamental idea behind the model of the Archaic Universe [3,4,8,9] is based on the hypothesis that there exists a phase prior to the big bang in which the geometry of the Universe is that of the four-dimensional surface of a hypersphere in the five-dimensional Euclidean space. The existence of a privileged axis in the five-dimensional space is hypothesized and, therefore, that of a privileged system of parallel circles (*i.e.*, three-dimensional spaces) on the hyperspherical surface. These circles are the intersections of the hyperspherical surface with the hyperplanes perpendicular to that axis. Each circle represents the Universe at a particular instant of the "time precursor", $\underline{x}_0$, that spans from the value $\underline{x}_0 = 0$ (corresponding to the position where the equator can be found) to the value $\underline{x}_0 = \pm r$ (where the poles are located). Here, the quantity $r$ is the radius of the hypersphere. The peculiarity of this system of parallel circles is specified by the assumption that the hypersphere is the site of virtual quantum fluctuations, which originate on the equator and ends on one of the several parallels. The complex of the fluctuations ending on $\underline{x}_0$ is assimilated to a thermostat at the absolute temperature $T$, according to the uncertainty principle written in the form $\underline{x}_0 = \hbar c/kT$, where $k$ is the Boltzmann's constant. The set of fluctuations constitutes a pre-vacuum. This pre-vacuum has free energy, does not contain matter and does not live in time, but is defined on the completely spatialized hyperspherical surface.

The hypersphere is therefore, in essence, the timeless seat of the pre-vacuum. What makes possible the transition of the Universe from this state to that, temporally ordered, of the ordinary vacuum plus the matter is the hypothesized instability of the pre-vacuum with respect to the formation of micro-events. In other words, one assumes the existence of a time interval $\theta_0$ such that, for $\underline{x}_0/c > \theta_0$, the free energy of the pre-vacuum can be converted into effective interactions between real elementary particles. The interval $\theta_0$ is identified with the time $t_e$ required for light to travel the classical radius of the electron ($t_e \approx 10^{-23}$ s). This time interval identifies the particle scale, namely, the scale on which matter appears granular because it is made up of micro-events of interaction between "elementary particles". By comparison, the interval $t_0 = r/c$ is estimated to be on the order of $10^{18}$ s, as in Ref. [8]. Here, the ratio between the two-time intervals surprisingly corresponds to the well-known Dirac number, $N_D \approx 10^{41}$. In summary, matter does not consist of elementary objects but instead is built with micro-events and the time scale of these micro-events is $\theta_0$. When the quantity $\underline{x}_0/c$ exceeds this time, the matter appears. The free energy of the pre-vacuum is then converted into matter and what remains is nothing but the ordinary vacuum.

This process of conversion into matter does not take place on the timeless realm of the hypersphere, which is eternally identical to itself. It instead connects the hypersphere to a different geometric structure: spacetime, which constitutes the appropriate environment to matter. Clearly, this connection between different geometric structures is not dynamic, since it is a connection between an a-temporal realm and the space-time domain, and consists of three geometric transformations. The first transformation is a Wick rotation, which turns the hypersphere into a hyperboloid. The second is the projection of the hyperboloid on the hyperplane tangent to it at any point P; this projection is conducted from the centre of the hyperboloid and is external to the five-dimensional light cone. These two first transformations induce on the hyperplane the Beltrami representation of a de Sitter space with radius *r*. The third transformation is the contraction of the Beltrami coordinates of particles by a $N_D$ scale factor that leaves *r* unchanged. The result is a spacetime inhabited by matter that is nucleating in the big bang due to the conversion process, as it appears at the observation point-event P. If the time axis in the hyperplane is coplanar to the privileged axis in the five-dimensional space, then P is a "fundamental" observer. All these various "private" spaces of the various "observers" P are connected to each other by de Sitter transformations. They are nothing but the different "points of view", on the same universe, of the different observers present at the big bang. The big bang is no longer a point-like and instantaneous event, a singularity, but takes place on the scale fixed by $\theta_0$. This is the geometric structure that constitutes the initial condition for the application of the gravitational equations under the hypothesis of the cosmological principle (guaranteed by the premises of the model). This application, in turn, leads to the spatially flat FRWL metric [3].

The Archaic Universe is thus the predecessor of the ordinary spacetime in an ontological, rather than chronological, sense. Hence the adjective "archaic", which refers to the meaning of the concept of Archè (basic, original) [4,10]. The Archaic Universe shapes the initial condition of the universe that we observe and therefore constitutes, in a certain sense, its original form.

To address the problem of the CPT-symmetry of the pre-vacuum, it is essential to understand that the hypersphere is a logical – and not chronological – antecedent of the big bang and spacetime. It is not a region with a Wick-rotated metric separated from the region with FRWL metric by some "border", as could be a Hartle-Hawking solution [9]. So, no physical signal can pass through such a non-existent border. The connection between the hypersphere and spacetime consists of a non-dynamic generative process represented by three geometric transformations. We must therefore turn our attention to these transformations.

Let us reconsider, in particular, the projection from the center O of the hyperboloid on the hyperplane tangent to the hyperboloid in P, with the segment OP lying outside the five-dimensional light cone with vertex in O. It is clear that for every projection of this type there exists a projection on a hyperplane parallel to that one considered, and it is tangent to the hyperboloid in the antipode P' of P with respect to O. For the sake of simplicity, we will assume that the spatial axes and the time axis of the two hyperplanes are parallel and equiverse in the five-dimensional space and assume that their origins are P and P', respectively. In this case, if we consider a point Q on the tangent hyperplane in P, the five-dimensional segment QO can be prolonged to the other hyperplane. This extension intersects the tangent hyperplane in P' in a point Q', whose spacetime coordinates in this hyperplane are exactly the opposite to those of Q in the tangent hyperplane in P. In other words, the two projections are conjugated both in parity and in time inversion.

If P, and therefore P', are images of points that were on the equator of the hypersphere, then the points Q and Q' are the images of points with positive and negative $x_0$ coordinates, respectively. It is not possible to move these points continuously on the hyperboloid until they coincide, without crossing the equator. These point-events are therefore and at all effects, "separated from the big bang", *i.e.,* separated from the region of the purely virtual processes $-\theta_0 < x_0 < +\theta_0$ at the two extremes of which nucleation occurs. They are causally-disconnected events in the ordinary sense, because they are children of the two twin big bangs downstream with $x_0 = +\theta_0$ and $x_0 = -\theta_0$ values, respectively. Nevertheless, they are connected by their common emergence from the pre-vacuum. In summary, we have two specular big bangs that produce two universes that are the one the PT = C mirror of the other, where the operators C, P, T represent respectively the charge conjugation, the spatial inversion and the temporal inversion and the equivalence symbol "=" has to be intended in the sense of the CPT invariance. A direct consequence is that the content of matter and antimatter of one of the two universes equals the content of antimatter and matter of the other one, respectively. If one of the two universes is dominated by ordinary matter, the other will be dominated by antimatter.

If one assumes that the sign of the charges is defined by the sign of $x_0$, *i.e.* by the fact that they appear at the time of the big bang in our universe or, alternatively, at the time of the anti-big bang in the anti-universe, one does indeed obtain the required separation *ab initio* of matter and antimatter (see footnote 1).

If both the temporal axes of the hyperplanes are equiverse to the $x_0$ axis in the five-dimensional space, then the energy released by the two big bangs is equal and opposite: one of the two universes has negative total energy. If one chooses an opposite orientation of the temporal axes in the two universes, then the energy of both universes is positive. In the first case, the change of sign of the energy is consistent with the already mentioned result concerning the temperatures discussed by Volovik [2].

## 2. **CPT symmetry in the archaic pre-vacuum**

---

1This is possible in a description of the co-emergence of internal quantum numbers and space-time coordinates in a single micro-event. In [11] it is hypothesized that a set of quaternionic units $V = \{1,i,j,k,-1,-i,-j,-k\}$ undergoes a separation in a retarded part $V^+$ and in an advanced part $V^-$. There are the two possibilities $V^+ = \{1,i,j,k\}$, $V^- = \{-1,-i,-j,-k\}$ and $V^+ = \{1,-i,-j,-k\}$, $V^- = \{-1,i,j,k\}$ corresponding respectively to the projection on a particle and antiparticle state. Only the retarded parts $V^+$ survive in the big bang. If the original fluctuations of the two signs of $x_0$ are separated by chirality, then there is particle nucleation for one sign and antiparticle nucleation for the other. This seems plausible if it is admitted that the particles originate from the same structural elements of space and it is observed that the two universes are of opposite parity.

In [1], Boyle, Finn and Turok start their analysis from a line element of the type $ds^2 = \eta_{ab} e^a e^b$, with $\eta_{ab} = diag\{-1,1,1,1\}$ and the tetradic 1-forms defined as $e_0 = -Nd\tau$, $e_i = h_{ij}^{1/2} dx^j$. The big bang occurs at $\tau = 0$. Then they define the tetrad at the time $\tau$ after the big bang so that it is the reverse of the tetrad at the corresponding instant of time before the big bang, on the same line $x^i = const$,

$$e_\mu^a(\tau, \boldsymbol{x}) = -e_\mu^a(-\tau, \boldsymbol{x}). \quad (1)$$

This choice is exactly the opposite of that adopted in our approach discussed in the previous section and it leads to a change of sign of the scale factor through the big bang. The resulting physical representation is, however, the same: it implies the cancellation of the singular modes of the scalar, vector and tensor perturbations at $\tau = 0$. From Eq. (1) it also follows that $e_\mu^a(0, \boldsymbol{x}) = 0$; the authors interpret the contraction of the tetrad at the big bang and its re-emergence "from the other side" as the creation of a universe-antiuniverse pair whose elements are PT-conjugated.

In addition, if one considers a FRWL background equipped with an isometry under time reversal $\tau \to -\tau$, then there is a preferred vacuum that respects the full isometry group (including CPT). In other words, by imposing a CPT Universe one constrains the vacuum. In this regard, these three authors consider a Weyl-invariant spinor field given by

$$\psi = a^{\frac{3}{2}} \Psi \quad (2)$$

where $\Psi$ is the spinor with positive-defined mass on a flat FRW background and the parameter $a$ is the background scale factor. The field in Eq. (2), which satisfies a Dirac equation of motion, can be expanded as follows

$$\sum_h \int \frac{d^3 \boldsymbol{k}}{(2\pi)^{3/2}} [a_0(\boldsymbol{k}, h)\psi(\boldsymbol{k}, h, x) + b_0^+(\boldsymbol{k}, h)\psi^c(\boldsymbol{k}, h, x)] \quad (3)$$

where the operators $a_0$ and $b_0^+$ are defined as

$$\begin{bmatrix} a_0(\boldsymbol{k}, h) \\ b_0^+(-\boldsymbol{k}, h) \end{bmatrix} = \begin{bmatrix} \cos\frac{\lambda(k)}{2} & \mp i sen\frac{\lambda(k)}{2} \\ \mp i sen\frac{\lambda(k)}{2} & \cos\frac{\lambda(k)}{2} \end{bmatrix} \begin{bmatrix} a_\pm(\boldsymbol{k}, h) \\ b_\pm^+(-\boldsymbol{k}, h) \end{bmatrix} \quad (4)$$

in such a way that they transform as $[CPT]a_0(\boldsymbol{k}, h)[CPT]^{-1} = -b_0(\boldsymbol{k}, -h)$ and $[CPT]b_0(\boldsymbol{k}, h)[CPT]^{-1} = -a_0(\boldsymbol{k}, -h)$. This implies that the corresponding vacuum, defined by $a_0|0_0\rangle = b_0|0_0\rangle$, is CPT-invariant.

The reader can easily grasp the similarities with our approach by comparing these brief references to paper [1] with the contents of the previous section. However, this comparison also immediately shows important differences. Our projective approach presents a symmetry which is lacking in more conventional cosmologies. In a pre-big bang phase there is a fundamental symmetry between the two hemispheres into which the infinite-temperature equator divides the five-dimensional hypersphere while after the big bang our Universe develops from one of these two hemispheres. Instead, the other hemisphere does not play any role and can be considered as a simple mathematical artefact as it corresponds to the choice of a negative sign (instead of positive) in the definition of the projective coefficient of the metric [12,13,14]. This second hemisphere can be regarded as a second archaic universe that is specular with our own Universe and is characterized by the archaic fluctuations exiting from the equator, which nucleate into particles that are the C-conjugated of the particles nucleating in our Universe. Thus, in the second hemisphere, an antimatter-dominated Universe would be created in such a way that the symmetry of the archaic pre-vacuum is maintained. These two "mutually specular universes" would be separated by the equator, and would therefore be causally unconnected, albeit contiguous. Their common origin would be just the archaic vacuum. The CPT symmetry can thus receive a new, more natural reading and explanation. In fact, while in BFT model the vacuum is bound by imposing a CPT universe, thanks to the *ad hoc* assumptions in Eq. (1) and (4), in our approach the CPT symmetry emerges naturally in virtue of the features and physical dynamics regarding the five-dimensional hypersphere and its corresponding projective relation with the private spacetimes.

As discussed in the previous section, a negative energy can be associated with the quantum fluctuations on the anti-hemisphere. Due to the symmetry of the initial state at infinite temperature, this negative energy exactly compensates the positive energy of fluctuations on the matter hemisphere, so enabling a *creatio ex nihilo*. However, for an inertial observer exiting from the anti-big bang in the anti-universe, which measures a growing cosmic time ($t \to -t$), the energy released at big bang, even in form of antimatter, is positive.

The results concerning the elimination of the singular components of the perturbations (scalar, vector or tensor) strongly depend on the elimination of even modes around the big bang [1], and therefore appear applicable to the scenario of the Archaic Universe. Although in this scenario the big bang is no longer a "boundary" between the two universes, but rather the common origin of both. It is then possible to speculate on perturbations originating from the fluctuations of the archaic vacuum at the nucleation. This idea has been explored in [15] (and in the references cited therein) with reference to the dark matter. Further research could lead to signatures of the archaic phase in the current Universe, useful for the validation of the proposed theoretical frame.

### 3. Archaic scenario and micro-events

In the Archaic Universe scenario, the elementary nucleations of matter are identified with the processes of reduction of the wavefunction that in turn are identified with quantum jumps (QJs). For the purposes of this article, we will refer to the specific model of QJ proposed in [7] that we will now propose again in a revised form. The first idea behind the model is that the nature of time is not that of a real variable but that of a complex variable. There is therefore a complex plane of time, in which each point ($\tau'$, $\tau''$) is identified by its real part $\tau'$ and its imaginary part $\tau''$. Each point of the plane is placed at a distance from the origin equal to $\rho = [(\tau')^2+(\tau'')^2]^{1/2}$. The structure of the plane is not affine, but instead linear. In particular, the origin (0,0) corresponds to a condition of complete timelessness in which the flow of events is completely arrested. The proper time axis of a particle is built with the points of the circumference $\rho = \theta_0$, in the sense that the oriented arcs counted on this circumference, starting from a point taken as reference, are in biunique correspondence with the instants of time proper of that particle. It should be noted that this definition does not involve any notion of spatial localization. There is a triple continuous infinity of temporal axes that can be associated with a free particle for each choice of the reference point on the circumference. The latter can in turn be spatially qualified in a triple continuous infinity of modes, each corresponding to a spatial position. However, the "spatial position" of the particle is not defined a priori, in agreement with the concept of delocalization of quantum entities. The extension to any number of particles, and therefore to configurational spaces, is immediate, but for our purposes it is sufficient to study the behavior of a single free particle.

The particle "travels" the circumference and this allows us to label it with a complex number which, in the simplest case of a free particle, is nothing but the de Broglie phase factor $\exp(iMc^2\tau/\hbar)$, where $M$ is the particle mass and $\tau$ the proper time (understood in the usual sense) passed from the reference instant. The complex number in mention identifies the end of the arc traveled, and is normalized to $\theta_0$, the radius of the circumference. Depending on the direction of the travel of the circumference there are two complex-conjugate numbers $\exp(+iMc^2\tau/\hbar)$ and $\exp(-iMc^2\tau/\hbar)$, which are both solutions of the equation:

$$-i\hbar\frac{\partial \Phi}{\partial \tau} = \pm Mc^2\Phi; \qquad (5)$$

The generic wavefunction $\Psi$ of the particle is a linear superposition of phase factors of this type, endowed with specific covariance properties with respect to the transformation of the spatial coordinates (scalar, spinor, etc.). A micro-event (QJ) is an instant, on the circumference $\rho = \theta_0$, from which a retarded function $\Psi$ and an advanced function $\Psi^*$ come out respectively e.g. in the anticlockwise and clockwise sense, respectively. These two functions are complex conjugate with respect to each other and carry positive and negative energy, respectively. If we consider two successive QJs in the history of a particle, they are causally connected in the sense that the retarded function of the first and the advanced one of the second QJ are respectively the ket and the bra between which the time evolution operator of the particle, in the interval between the two instants, is "sandwiched". The result is the transition amplitude between the two events.

Each event, considered individually, is represented by the corresponding projector $|\Psi\rangle\langle\Psi|$ on the outgoing wave function. The particle thus becomes the localization, in the time domain $\tau$ (subset of the complex time domain) of the packet of physical quantities associated with the wavefunction $\Psi$. We can hypothesize that the time domain $\tau$ and the QJs emerge from an antecedent state, represented by a pre-vacuum fluctuating around the a-temporal condition $\rho = 0$. Assuming an exponential probability of fluctuations of the type:

$$P = \exp\left(-\frac{\rho}{\theta_0}\right) \qquad (6)$$

it can be expressed as the square modulus of the function $\Lambda(\rho)$, solution of the equation

$$-i\hbar\frac{\partial \Lambda}{\partial(i\rho)} = \frac{\hbar\Lambda}{2\theta_0}, \qquad (7)$$

which constitutes the second equation of the revised model considered by us. The connection with the archaic scenario then consists of the following hypothesis:

$$\rho \leq \frac{|x_0|}{c} = \frac{\hbar}{kT} \qquad (8)$$

where $k$ is the Boltzmann constant and $T$ is the (absolute) temperature of the thermostat consisting of the fluctuations originated in $x_0 = 0$ and ending in the generic point $x_0$. As a result of Eq. (8), the distribution in Eq. (6) is truncated to the maximum value of $\rho$ given by $|x_0|/c$. Therefore, QJs cannot occur for $|x_0|/c < \theta_0$: this is the condition of the archaic pre-vacuum. QJs become possible as soon as $|x_0|/c$ exceeds $\theta_0$; in fact, the circumference $\rho = \theta_0$ is now included in the domain accessible to the fluctuations and the breaking of symmetry that leads from the false vacuum $\rho = 0$ to the true vacuum $\rho = \theta_0$ can therefore take place. In essence, the pre-vacuum enters the time domain $\rho = \theta_0$ which becomes its new stable vacuum. And it is in this vacuum that QJs nucleate. This nucleation is the big bang, and since the minimum extension of the $\tau$ domain accessible to fluctuations is given by the length of the circumference which is $\approx \theta_0$, is also the temporal thickness of the big bang. The big bang therefore occurs in a finite time interval, it is not an instant-singularity.

The cosmogenesis is therefore governed by the ratio of $t_0$ (in turn connected to the cosmological constant) and $\theta_0$. This ratio, on the order of the Dirac number, defines the amount of information associated with the localization of a particle in the time domain, an amount that is infinite. If the initial stock of information is assumed to be finite, the particles will be nucleated in a finite number, albeit enormous. Once this nucleation is terminated (in the sense of $x_0$) there will be two universes in the ordinary sense of the term, with angular or $\tau$ times. The radial time $\rho$ will only play a role in micro-interactions, and it will be hidden within them.

The first universe will originate from the nucleation at $x_0/c = +\theta_0$, the other from the nucleation at $x_0/c = -\theta_0$. As we have seen, these two universes are one the mirror of the other in the sense of the conjugation of charge. If we assume that the original QJs manifested in the big bang are asymmetric, in the sense that they lead to the emission of the only retarded wavefunction for $x_0 > 0$ and the only advanced wavefunction for $x_0 < 0$, from the Feynman-Stuekelberg interpretation the *ab initio* separation between matter and antimatter automatically follows, under the assumptions illustrated in footnote 1.

It can be noted that these hypotheses explain in a simple way two assumptions that in the original formulation of the model of the Archaic Universe were simply postulated: the Wick rotation (*i.e.,* the passage from $x_0$ to $\tau$) and the scale reduction of a factor equal to the Dirac number, which is reproduced through the finite temporal extension $\approx \theta_0$ of each of the two nucleations. Since the model includes the cosmological principle, the third assumption concerning the dependence of the scale factor on the cosmic time after the big bang is a direct consequence of the application of the relativistic formalism in each of the two universes, after the nucleation.

## 4. Conclusive remarks

Our proposal permits to explain the dominance of ordinary matter without introducing special initial conditions, at the same time preserving the greatest symmetry of the initial state in the sense that is not requested to postulate the CPT symmetry *ad hoc*. In other words, our approach allows us to explain the discrepancy matter-antimatter by obtaining CPT symmetry as a direct consequence of the features of the two hemispheres of the five-dimensional hypersphere and their corresponding vacuum fluctuations.

Moreover, if the mechanism based on which ordinary matter and antimatter would seem to separate along the equator remains enigmatic, it appears plausible to presume that the particles acquire their own charges only at the time of the big bang, when they become real and therefore capable of real interaction. If one assumes that the sign of the charges is defined by the hemisphere in which the particle becomes real (*i.e.,* by the fact that it appears at the time of the big bang in our Universe or, alternatively, at the time of the anti-big bang in the anti-universe) one does indeed obtain the required separation *ab initio* of matter and antimatter. The sign of the charge would in other words be defined by the direction of the timeline emerging from the equator along which the particle materialized, a result which could in some way be connected with the CPT theorem.

On the other hand, also the fact – invoked by Volovik in [2] – that the big bang emerges as the bifurcation point of the second order quantum transition from the Euclidean to the Minkowski signature, at which the symmetry between the spacetime and anti-spacetime is spontaneously broken, receives here a natural reinterpretation as a passage from the (locally Euclidean) metric of the five-dimensional hypersphere to the (locally Minkowski) metric of private spacetimes. If in the picture proposed by Volovik the initial stage of the evolution of our Universe after the big bang can be characterized by the negative temperature, this is due to the fact that a negative energy can be associated with quantum fluctuations on the anti-hemisphere and, in virtue of the symmetry of the initial state at infinite temperature, this negative energy exactly compensates the positive energy of fluctuations on the matter hemisphere. The features of the fluctuations of the vacuum characterizing the two hemispheres of the five-dimensional hypersphere determine the nature of the big bang as a bifurcation point where a spontaneous symmetry breaking between matter and antimatter occurs.

In this framework, the CPT symmetry is a necessary consequence of the archaic vacuum and its nucleation process, differently from the BMT model where CPT was assumed a priori. We notice that this approach through a de Sitter universe rebuilds the Hartle-Hawking conditions avoiding any conflict with the eternal de Sitter and the FRW$\Lambda$ cosmological model [9,17]. The big bang, at all effects, becomes in a certain sense nothing but the "appearance" of classical information.

This approach can be extended also to weaker conditions that involve the PT symmetry only, for which new classes of complex Hamiltonians can admit real and positive spectra like those of the self-adjoint (Hermitian) Hamiltonians [16]. In fact, in this model one can already insert an imaginary time component in the de Broglie phase factor. This is equivalent to a rescaling of the phase factor norm and to a shift of an additional quantity $\rho$ from the origin (0,0) different from the initial chosen value $\theta_0$ allowing, from Eq. (7), the extension of this model to a wider class of real and positive value non-Hermitian Hamiltonians.

## Acknowledgements

FT acknowledges ZKM and Peter Weibel for the financial support.